\documentclass[epj]{svjour}
\usepackage{graphics}
\begin{document}
\title{Recent Charm from CLEO}
\author{Yongsheng Gao\inst{1}}                     
\institute{Physics Department, Southern Methodist University,
           Dallas, TX 75275, USA}
\date{Received: \today / Revised version: \today}
%
\abstract{
We present some recent results in rare charm decays
from CLEO Collaboration. The data used were collected
by the CLEO II and III detectors at the Cornell 
Electron Storage Ring (CESR). A brief future outlook for 
the CESR-c/CLEO-c program is also presented.
} 
%
\maketitle
\section{Introduction}
\label{intro}
Charm sector is a good hunting ground for possible new physics
beyond the Standard Model (SM). Because the SM processes are highly 
suppressed, the opportunity for evidence of new physics to emerge 
becomes enhanced in charm relative to the other flavors~\cite{burdman}.

\section{Cabibbo-Suppressed $D^{+}$ Decays }
\label{sec:1}
$D^{+} \to \pi^{+}\pi^{0}$ and $K^{+}K_{s}^{0}$ are single 
Cabibbo-Suppressed decays and $D^{+} \to K^{+}\pi^{0}$ is a doubly 
Cabibbo-Suppressed decay. One of the main interest in charm physics 
has been the $D^{0} - \bar{D^{0}}$ mixing. To unravel any non SM 
contributions to $D^{0} - \bar{D^{0}}$ mixing, we need to understand 
the SU(3) symmetry breaking effects. These Cabibbo-Suppressed decays 
will provide useful estimation of the SU(3) violating effects in the 
$D$ system.

Given the large uncertainties in absolute $D^{+}$ branching fractions, 
we use $D^{+} \to K^{-}\pi^{+}\pi^{+}$ and $K_{s}^{0}\pi^{+}$ as 
normalization modes and measure the ratios of these Cabibbo-Suppressed 
decays relative to these normalization modes.

We perform maximum likelihood fit to the selected data sample and extract 
the signal from unbinned maximum likelihood fits. The signal yields are 
summarized in Table~\ref{tab:CSDplus}.
They translate into the following ratios of branching fraction
measurements: 
$\frac{Br(D^{+} \to \pi^{+}\pi^{0})}
      {Br(D^{+}\to K^{-}\pi^{+}\pi^{+})}$ = 0.0144 $\pm$ 0.0019 $\pm$ 0.0010,
$\frac{Br(D^{+} \to K^{+}K_{s})}
      {Br(D^{+}\to \pi^{+}K_{s})}$        = 0.1892 $\pm$ 0.0155 $\pm$ 0.0073,
$\frac{Br(D^{+} \to K^{+}\pi^{0})}
      {Br(D^{+}\to K^{-}\pi^{+}\pi^{+})}$ = 0.0029 $\pm$ 0.0018 $\pm$ 0.0009.

\begin{table}
\caption{Yields from the maximum likelihood fit with statistical errors
         and reconstruction efficiencies.}
\label{tab:CSDplus}      
\begin{tabular}{ccc}
\hline\noalign{\smallskip}
Mode & Yield & Efficiency  \\
\noalign{\smallskip}\hline\noalign{\smallskip}
$\pi^{+}\pi^{0}$      &   171.3$\pm$22.1  & (6.20$\pm$0.11)\% \\
$K^{+}K_{s}$          &   277.7$\pm$20.8  & (4.94$\pm$0.23)\% \\
$K^{+}\pi^{0}$        &    34.3$\pm$20.9  & (6.08$\pm$0.22)\% \\ \hline
$K^{-}\pi^{+}\pi^{+}$ & 12898.0$\pm$156.6 & (6.74$\pm$0.12)\% \\
$\pi^{+}K_{s}$        &  1434.7$\pm$48.0  & (4.83$\pm$0.23)\% \\
\noalign{\smallskip}\hline
\end{tabular}
\end{table}

Using PDG~\cite{PDG} values of $Br(D^{+}\to K^{-}\pi^{+}\pi^{+})$ = 
(9.1 $\pm$ 0.6)\% and $Br(D^{+}\to \pi^{+}\bar{K^{0}})$ = 
(2.77 $\pm$ 0.18)\%, we obtain the following branching fraction 
measurements:
$Br(D^{+}\to \pi^{+}\pi^{0})$  = (1.31 $\pm$ 0.17 $\pm$ 0.09 $\pm$ 0.09)
                                 $\times$ 10$^{-3}$,
$Br(D^{+}\to K^{+}\bar{K^{0}})$ = (5.24 $\pm$ 0.43 $\pm$ 0.20 $\pm$ 0.34)
                                 $\times$ 10$^{-3}$,
$Br(D^{+}\to K^{+}\pi^{0})$     = (2.64 $\pm$ 1.64 $\pm$ 0.82 $\pm$ 0.17)
                                 $\times$ 10$^{-4}$ and
$Br(D^{+}\to K^{+}\pi^{0})$ $<$ 4.2 $\times$ 10$^{-4}$ at 90\% C.L.
The first error is statistical and the second error is systematic. The third
error in the measurements is due to the uncertainty in the normalization 
braching fractions.
For $\pi^{+}\pi^{0}$ and $K^{+}K_{s}$ decays, these numbers are the best 
single measurements. For $K^{+}\pi^{0}$. These numbers give the first limit 
on this decay mode.

We also measured the ratio R1~\cite{chau} which is expected to be 1 in 
the limit of SU(3) symmetry. Our measurement of 1.84$\pm$0.38 is higher than 
the theoretical predition that the SU(3) symmetry breaking effects are at 
about 30\% level.

It is believed that in the $D$ system the interference between external and
internal decay amplitudes is destructive. 
Our measurement of R2~\cite{chau} of 2.03$\pm$0.32 indicates this is indeed 
the case.

The details of the analysis can be found in~\cite{clns03-1832}

\section{Dalitz Analysis of $D^{0} \to \pi^{+}\pi^{-}\pi^{0}$}
\label{sec:2}
Three-body decays provide excellent opportunities to study the interference
among the intermediate state resonances, allowing the measurements of both 
the amplitudes and phases of the intermediate states.
For example, in the Dalitz analysis of $D^{+} \to \pi^{+}\pi^{-}\pi^{+}$, 
E791 reported significant evidence for a borad neutral scalar 
resonance, the $\sigma$(500)~\cite{e791}.

\begin{figure}
\resizebox{0.45\textwidth}{!}{%
  \includegraphics{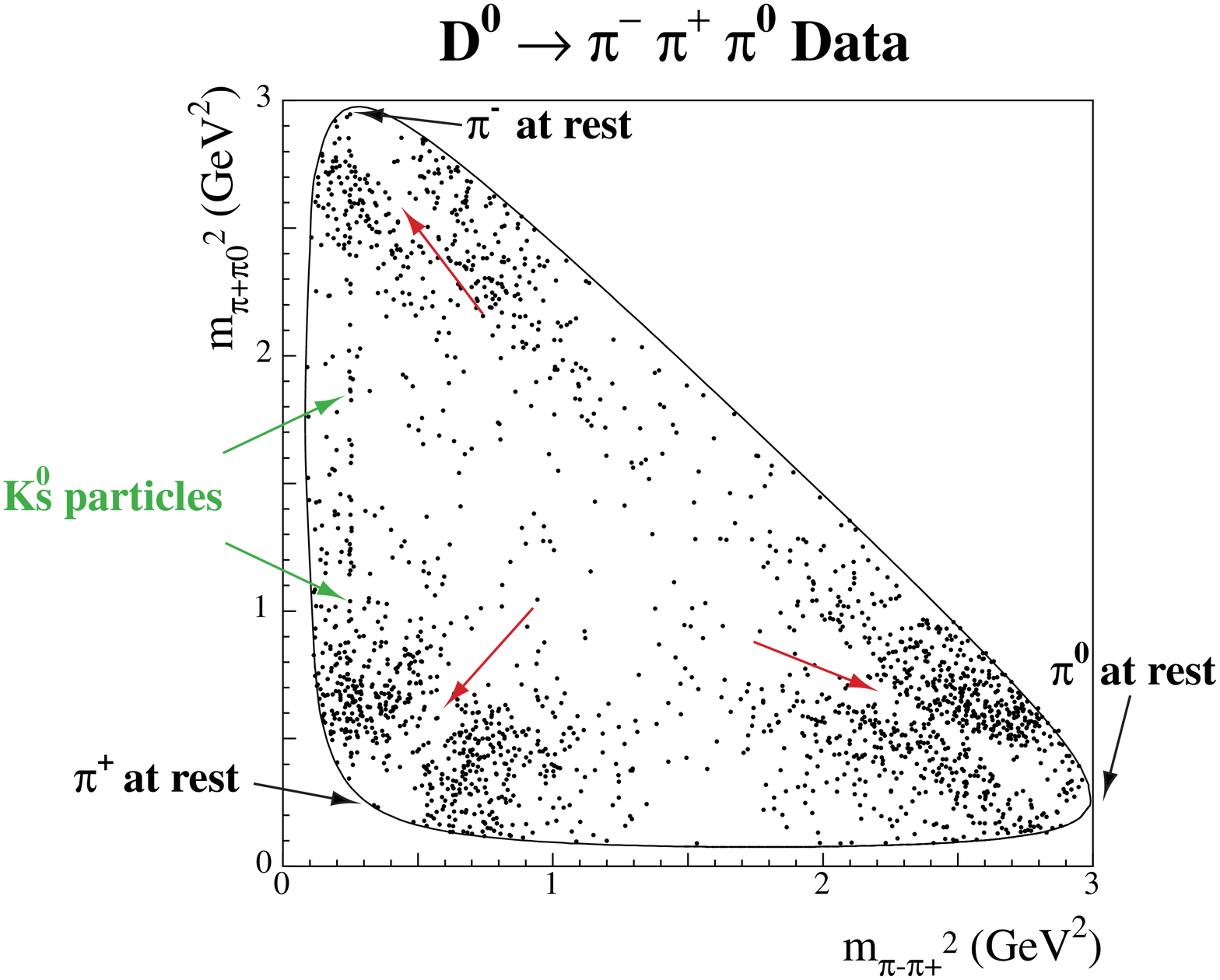}
}
\caption{Dalitz plot for $D^{0}\to\pi^{+}\pi^{-}\pi^{0}$ from CLEO data.}
\label{fig:dalitz}       
\end{figure}

We performed Dalitz analysis of $D^{0} \to \pi^{+}\pi^{-}\pi^{0}$ at CLEO. 
About 80\% of the events entering the Dalitz plot in Figure~\ref{fig:dalitz}
are signal events. 
We then extact the amplitude and phase of each component. The results 
are summaried in Table~\ref{tab:dalitz}.

\begin{table}
\caption{Dalitz analysis fit results.}
\label{tab:dalitz}      
\begin{tabular}{cccc}
\hline\noalign{\smallskip}
Resonance & Amplitude & Phase ($^{0}$) & Fraction (\%)  \\
\noalign{\smallskip}\hline\noalign{\smallskip}
$\rho^{+}$    & 1.0 (fixed)             & 0.0 (fixed)     & 76.5$\pm$1.8$\pm$4.8
 \\
$\rho^{0}$    & 0.56$\pm$0.02$\pm$0.07  & 10$\pm$3$\pm$3  & 23.9$\pm$1.8$\pm$4.6
 \\
$\rho^{-}$    & 0.65$\pm$0.03$\pm$0.04  & $-4\pm$3$\pm$4  & 32.3$\pm$2.1$\pm$2.2
 \\
 non res.     & 1.03$\pm$0.17$\pm$0.31  & 77$\pm$8$\pm$11 &  2.7$\pm$0.9$\pm$1.7
 \\
\noalign{\smallskip}\hline
\end{tabular}
\end{table}

Including a scalar $\sigma (500)$ does not result in a significantly 
improved likelihood and yielded fit fraction is consistent with zero.

It is also interesting to mention that we do not see more massive $\rho$
mesons like $\rho^{+}(1700)$ either.

The details of the analysis can be found in~\cite{cleo-conf-03-03}

\section{First Observation of $D^0 \to K^0_S \eta \pi^0$}
\label{sec:3}
This work was motivated by BaBar and CLEO's recent Dalitz analyses 
results of $D^{0} \to K^{0}_{s} K^{+}K^{-}$ and $K^{0}_{s} \pi^{+}\pi^{-}$. 
As a$_{0}$(980) to $\eta \pi^{0}$ is the dominating decay mode, we expect to 
observe $D^0 \to K^0_S \eta \pi^0$ as well.

\begin{figure}
\resizebox{0.45\textwidth}{!}{%
  \includegraphics{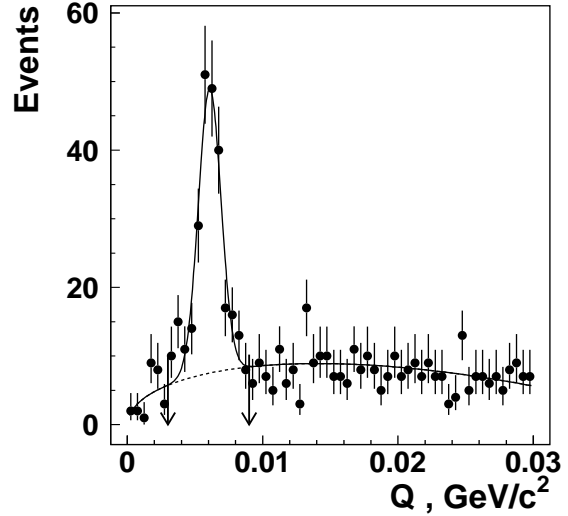}
}
\caption{The Q distribution for the decay
            $D^{*+} \to D^0(K^0_S \eta \pi^0) \pi^+$.}
\label{fig:ksEtaPi0}       
\end{figure}

To study this decay we use decay channels: $K^0_S \to \pi^+\pi^-$, 
$\eta \to \gamma \gamma$, $\pi^0 \to \gamma \gamma$.
The decay $D^0 \to K^0_S \pi^0$ 
is used for systematic cross-checks and normalization.
Using the energy release variable $Q=M(D^{*+})-M(D^0)-m_{\pi^+}$,
we observe a clean signal in the $Q$ distribution as shown in 
Figure~\ref{fig:ksEtaPi0}. This is the first observation of the decay 
mode of $D^0 \to K^0_S \eta \pi^0$.
The preliminary measured the ratio of branchings fractions is:
\[ 
     R = \frac{BR(D^0 \to K^0_S \eta \pi^0)} {BR(D^0 \to K^0_S \pi^0)}
       = 0.38 \pm 0.07_{stat.} \pm 0.05_{syst.}.
\]

We are performing Dalitz plot analysis for the decay 
$D^0 \to K^0_S \eta \pi^0$. Final results will be ready in the near
future.

\section{First Search for Flavor Changing Neutral Current Decay 
         $D^0 \to \gamma \gamma$}
\label{sec:4}
Standard Model (SM) predicts the rate for the $D^0 \to \gamma \gamma$ decay
of $\sim 10^{-8}$ or less~\cite{fajfer,burdman}. Non SM extension, i.e.
gluino exchange in SUSY, might enhance this rate by two orders of magnitude.
A measurement of this decay mode is a good test of new physics beyond the SM.

For this analysis \cite{Gao} we use CLEO~II \& II.V data sample.
Combinatoric background is suppressed by the tagging process 
$D^{*+}\to D^0\pi^+$ using the energy release variable 
$Q=M(D^{*+})-M(D^0)-m_{\pi^+}$.

\begin{figure}
\resizebox{0.45\textwidth}{!}{%
  \includegraphics{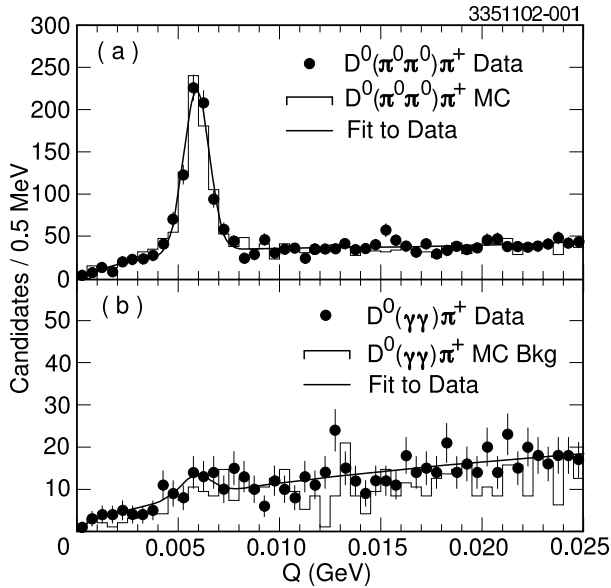}
}
\caption{Energy release in the decay $D^{*+}\to D^0\pi+$
         for (a) $D^0\to \pi^0 \pi^0$; (b) $D^0\to \gamma \gamma$.}
\label{fig:d0gg}       
\end{figure}

Figure~\ref{fig:d0gg} shows the $Q$ distributions for
$D^{*+} \to D^{0}\pi^{+}$ candidates where $D^{0} \to \pi^{0}\pi^{0}$ and
$D^{0} \to \gamma\gamma$. The circles with error bars are CLEO data
which are fit using a binned likelihood fit to a Gaussian function with 
expected mean and width determined from signal Monte Carlo simulation, 
on top of a threshold background function.
For $D^{*+}\to D^{0}\pi^{+}$ where $D^{0} \to \pi^{0}\pi^{0}$, 
$628.0 \pm 31.8$ signal events of $D^{0} \to \pi^{0}\pi^{0}$ are observed. 
The signal and background levels found in the data are in good agreement 
with those obtained from Monte Carlo simulations.
For $D^{*+}\to D^{0}\pi^{+}$ where $D^{0} \to \gamma\gamma$, no significant
enhancement is observed in the signal region. The signal yield of 
$D^{0} \to \gamma\gamma$ from the fit is $19.2 \pm 9.3$ events.
From Monte Carlo simulations, 
the relative efficiency for $D^{0} \to \gamma\gamma$ and
$D^{0} \to \pi^{0}\pi^{0}$ is determined to be:
$\epsilon(\gamma\gamma )/\epsilon(\pi^{0}\pi^{0})$ = 1.58 $\pm$ 0.05.
We set upper limits: 
     \[ B(D^0\to \gamma \gamma) / B(D^0\to \pi^0 \pi^0) < 0.033 \] and 
     \[B(D^0\to \gamma \gamma) <2.9\times10^{-5}~~@~~90\% C.L. \]

This is the first search for the decay mode of $D^0\to \gamma \gamma$.
The details of the analysis can be found in~\cite{Gao}

\section{CLEO-c and CESR-c project news}
\label{sec:5}

In early 2003, the NSF  approved a five year program of charm physics 
called CESR-c and CLEO-c\cite{CLEOc-CESRc}.
From 2003 to 2006, the Cornell Electron Storage Ring (CESR) accelerator
will operate at center of mass energies at 4140, 3770 and 3100MeV with the
expected luminosity of a few 10$^{32}$ cm$^{-2}$s$^{-1}$.

One may wonder what physics can CLEO-c offer in the next few years when
we expect lots of data from Tevatron and B factories. Especially BaBar and 
Belle will have a few hundred fb$^{-1}$ of data which contains lots of 
charm decays.

Running at Charm threshold with $D$ tagging provides an extremly powerful
and ``background free'' environment that allow some very precise measurements 
which are not possible at other environments.

For example, recently there have been exciting progress in Lattice QCD which 
will be able to calculate with accuracies of 1 to 2 percent. The CLEO-c decay 
constant and  form factor measurements with similar precision will provide a 
golden and timely test of Lattice QCD.

Large uncertainties in the current and future measurements of CKM matix
elements come from our poor measurements of the absolute charm branching
fractions. Again, CLEO-c will be the ideal place to provide these crucial
information in a timely fashion.

There are also potential to observe new forms of matter: Glueballs and hybrids.
As we all know, glueballs have been sighted too many times without confirmation.
With 1 billion $J/\psi$ data, CLEO-c can either find it or debunk it.

The CESR-c/CLEO-c program can be briefly listed as follow:
\begin{itemize}
\vspace{4mm} \item
     {\bf 2003 $-$ 2004 Act I: $\psi (3770)$ --- 3~fb$^{-1}$; 
          30M events, 6M {\em tagged} D
          (310 times MARK III).}
\vspace{4mm} \item 
     {\bf 2004 $-$ 2005 Act II: $\sqrt{s} \sim 4100~MeV$ --- 3~fb$^{-1}$; 
          1.5M $D_s \bar D_s$, 0.3M {\em tagged} $D_s$  
          (480 $\times$ MARK III).}
\vspace{4mm} \item 
     {\bf 2005 $-$ 2006 Act III: $\psi (3100)$ --- 1~fb$^{-1}$; 
          1~Billion $J/\psi$ 
          (170 times MARK III, 20 times BES II).}
\end{itemize}
This statistics is required for precise measurement of branching ratios,
decay constants and other SM parameters in the charm sector.

\end{document}